\documentclass[12pt]{article}

\usepackage{pstricks}
\usepackage{graphicx}
\usepackage[latin2]{inputenc}
\usepackage{epic}
\usepackage{latexsym}
\usepackage{epsf}
\usepackage{epsfig}

\newcommand{\eref}[1]{eq.\ (\ref{e.#1})}
\newcommand{\erefn}[1]{(\ref{e.#1})}

\newcommand{\cref}[1]{Chapter \ref{c.#1}}

\def\nn{\nonumber \\}
\newcommand{\nl}{& \nonumber \\ &}

\def\simgt{\stackrel{>}{{}_\sim}}
\def\beq{\begin{equation}}
\def\eeq{\end{equation}}
\def\bea{\begin{eqnarray}}
\def\eea{\end{eqnarray}}
\def\ba{\begin{array}}
\def\ea{\end{array}}
\def\bi{\begin{itemize}}
\def\ei{\end{itemize}}
\def\be{\begin{enumerate}}
\def\ee{\end{enumerate}}
\def\beq{\begin{equation}}
\def\eeq{\end{equation}}
\def\bc{\begin{center}}
\def\ec{\end{center}}

\def\pa{\partial}

\def\co{{\mathcal O}}

\def\t{\tilde}
\def\ov{\overline}

\def\ds{\displaystyle}

\def\s{\sigma}
\def\t{\tau}

\def\ads{${\rm AdS}_4 \,$}

\begin{document}

\pagestyle{empty}
\begin{flushright}
                   IFT-29/2004, KAIST-TH 2004/19
\end{flushright}
\vskip 2cm

\begin{center}
{\huge Stability of flux compactifications and the pattern of
  supersymmetry breaking}
\vspace*{5mm} \vspace*{1cm}
\end{center}
\vspace*{5mm} \noindent
\vskip 0.5cm
\centerline{\bf K. Choi ${}^a$, A. Falkowski ${}^b$, H. P. Nilles ${}^c$,
M. Olechowski ${}^b$, S. Pokorski ${}^b$}
\vskip 1cm
\centerline{\em ${{}^{a}}$  Department of Physics, Korea Advanced Institute of Science and Technology}
\centerline{\em Daejeon 305-701, Korea}
\vskip 5mm
\centerline{\em ${{}^{b}}$ Institute of Theoretical Physics, Warsaw
  University}
\centerline{\em ul.\ Ho\.za 69, PL-00 681 Warsaw, Poland}
\vskip 5mm
\centerline{\em ${{}^{c}}$ Physikalisches Institut, Universit\"at Bonn}
\centerline{\em Nussallee 12, D-53115 Bonn, Germany}
\vskip2cm

\centerline{\bf Abstract}
\vskip .3cm
We extend the KKLT \cite{kakali} approach to moduli 
stabilization by including the
dilaton and the complex structure moduli into the effective
supergravity theory. Decoupling of the dilaton is neither always
possible nor necessary for the existence of stable minima with zero
(or positive) cosmological constant. The pattern of supersymmetry
breaking can be much richer than in the decoupling scenario of KKLT.

\vskip .3cm


\newpage
\pagestyle{plain}
\section{Introduction}
One of the central questions in superstring theory is the
stabilization of moduli. In general,
the moduli fields include the dilaton $S$, the K\"{a}hler moduli $T_i$
and the complex structure
moduli $Z_i$. Early attempts in the heterotic string theory tried to
fix  the dilaton $S$ with a combination of gaugino condensation and nontrivial
flux of the 3-form field strength $H$ \cite{deibni}. 
The stabilization of $S$ turned out to be inherently connected to supersymmetry breakdown.
More recently it was observed \cite{gikapo}, that in the framework of 
Type II B theory one could fix complex
structure moduli as well as the dilaton with a combination of 3-form
fluxes of the field strengths $F_3$ and $H_3$ \cite{fluxes}, even in the absence of supersymmetry breakdown.
These attempts were only partially successful, as the moduli directions $T_i$ remained flat. 
Without a stabilization of the remaining moduli, 
it is difficult to draw specific conclusions about properties of the theories
 such as soft supersymmetry breaking terms.

In a more recent paper \cite{kakali} (KKLT) a combination 
of 3-form flux (in the Type II
B theory) and gaugino condensation was argued to lead to complete
stabilization of all moduli. The analysis in KKLT is done in the
framework of a low energy supergravity approximation. They assume that
the dilaton and the complex structure moduli (CSM), if present, have
been fixed by $H_3$ and $F_3$ fluxes and concentrate on an effective
theory for the volume modulus $T$ (representing the K\"ahler moduli
$T_i$). This decoupling procedure is self-consistent if the
masses for $S$ and CSM are much larger than the mass of
$T$. Non-perturbative corrections to the superpotential are then used
to stabilize $T$. In this analysis the ground state of the theory
had a large negative vacuum energy (AdS space) and preserved
supersymmetry. To reach an acceptable potential KKLT proposed a
so-called uplifting mechanism that breaks supersymmetry (in a local
minimum) and allows a fine tuning of the cosmological constant to a
desired value (e.g. de Sitter space). 


The purpose of the present paper is twofold. First, we include into
the effective supergravity theory the dilaton and the CSM.\footnote{A similar extension of the KKLT scenario has been considered in ref. \cite{bral}.} We compare
in a rather model independent way the results of such an extended
analysis with the KKLT results obtained under the decoupling
assumption. Secondly, we carefully analyze the pattern of
supersymmetry breaking in our extended framework. 
We shall see that a meaningful statement about the soft supersymmetry
breaking terms is only possible once all the moduli are stabilized.
The soft terms show an unexpectedly rich structure already in the 
(decoupled) KKLT limit and even more so in the general set-up.

We adopt the same assumptions for the supergravity analysis as in KKLT. 
In particular, we assume that the expectation values of $T$ and $S$ are large. In the region of large $S$ and $T$, there are simple formulae for the defining functions of supergravity: the K\"{a}hler potential $K$, the gauge kinetic function $f$ and the superpotential $W$. 
We concentrate on two specific schemes: $(D3/7)$, the original KKLT
scenario in type II B theory with matter on $D3$  \cite{kakali}  and/or $D7$ branes \cite{bukaqu}
and $(H)$, a heterotic theory with flux stabilization on a non-Calabi-Yau manifold \cite{nonCY} and gaugino condensation \cite{nilles}.

Our results can be summarized as follows (all within the approximation
stated above):
\begin{itemize}
\item
In the first step (flux compactification without gaugino condensation)
there are stable minima but there remains an unstabilized modulus 
(the flat direction $T$ in the Type II B case or the runaway direction $S$ in  the heterotic case),
\item
After inclusion of  gaugino condensation the existence of local 
(supersymmetric) minima in the
large $S$, $T$ region is not guaranteed in all cases.
It depends strongly on the model under consideration. In models
without complex structure moduli, there do not exist stable
supersymmetric AdS minima in the large $S$, $T$ region.
We find, with a quite general uplifting potential that for D3/7 a
stable minimum does not exist also after uplifting. Thus, in D3/7
models without CSM the KKLT procedure of decoupling $S$ is
inconsistent. For the heterotic model, a stable minimum does exist
after uplifting.
\item
In models with CSM, we derive the conditions on the (effective)
superpotential that allow for the existence of the supersymmetric AdS
vacuum, with all moduli stabilized. These conditions are restrictive
but more general than the decoupling limit of KKLT. Therefore, the
KKLT mechanism of stabilization works for a more general class of
models than those in which both CSM and $S$ can be decoupled.
The same conditions are sufficient for the existence of a stable local
minimum after uplifting.
\item
For the effective superpotential  satisfying the conditions allowing
for a stable minima, we discuss the supersymmetry breaking parameters
$F_S$ and $F_T$. We find that several other options than the KKLT
limit  $F_S \ll F_T \ll m_{3/2}$ are possible, with potentially
interesting consequences for phenomenology.
\end{itemize}

We stress here again that these results are obtained within the
supergravity approximation and the assumptions concerning $K, f $ and
$W$ as given explicitly in section 2, where we set up our notation and
review the mechanism of KKLT. Section 3 is devoted to an analysis
of the simplest case, with just the dilaton $S$, one K\"ahler modulus
$T$ and no complex structure moduli $Z_i$.
In section 4 we include complex structure moduli in a rather model
independent way. Section 5 contains an analysis of supersymmetry
breaking soft terms in the uplifted minima and the consequences for
realistic model building. Conclusions and outlook follow in section 6.
In the present paper we shall try to  avoid the presentation of
detailed technicalities of the calculations. These will be relegated
to a future publication where  also more examples are presented.

\section{Supergravity description of flux compactifications}

\subsection{Notation and conventions}

We concentrate on the case with the dilaton $S$, a  K\"{a}hler modulus
$T$ and complex structure moduli $Z_i$. Matter superfields are denoted
by $Q$. We assume to be in a region of large $S$ and $T$. Let us start
with the D3/7-system \cite{ibanez,louis,luest}. 
The K\"{a}hler potential is assumed to be
\begin{equation}
\label{kaehler-IIA}
K= - \log (S + \ov S - |Q_7|^2   ) - 3 \log (T + \ov T - |Q_3 |^2)
+ \tilde{K}(Z_i, \ov Z_i)
\end{equation}
where $Q_3$, $Q_7$ denote matter multiplets on the D3, D7 branes,
respectively. The gauge kinetic functions are
\begin{equation}
\label{gauge-IIA}
f_3 = S \hskip0.8cm {\rm or} \hskip0.8cm f_7 = T
\end{equation}
for gauge bosons on the D3 or D7 branes.
The superpotential is given by
\begin{equation}
W = W(S, Z_i) + C \exp (-aT) + W(Q_3, Q_7),
\end{equation}
where $C$ and $a$ are constants. The term $C\exp(-aT)$ represents
nonperturbative effects as explained in KKLT. When analyzing the
potential we look for minima where $Q_7$ and $Q_3$ scalars (and
therefore $W(Q_i)$ as well) do not receive nontrivial vacuum
expectation values. These minima coincide in the two cases $D3$ and
$D7$, while the consequences for the soft supersymmetry breaking terms
might differ.

In the heterotic case
$(H)$ we have
\begin{eqnarray}
\label{kaehler-het}
K_H &=& - \log (S + \ov S) - 3 \log (T + \ov T - | Q_H |^2)
+ K(Z_i, \ov Z_i) \nonumber\\
f_H &=&  S \\
W & = &  W(T,Z_i) + C \exp (-aS) + W(Q_H), \nonumber
\end{eqnarray}
where $C \exp(-aS)$ comes from gaugino condensation in the hidden
sector and $W(T,Z_i)$ comes from nontrivial fluxes on
specific non-Calabi-Yau manifolds \cite{nonCY}. 
As an example we consider here the
compactification on half-flat manifolds as given in ref. \cite{gulumi}.

\subsection{The proposal of KKLT}

The authors of ref. \cite{kakali} consider the D3/7 case and assume that the
dilaton and complex structure moduli are heavy compared to $T$, such that
they can be integrated out leaving a constant contribution $W_0$
to the effective low-energy superpotential $W_{\rm eff} = W_0 + C \exp
(-aT)$. This leads to a potential with a
supersymmetric AdS minimum in which the (last) modulus $T$ is fixed as
well. To this end, they add an uplifting
potential $\Delta V(T,\ov T)$ that breaks supersymmetry and allows a
local minimum at a (fine tuned)
small positive vacuum energy. The same proposal can be realized in the
heterotic case where $T$, $Z_i$ are
integrated out and $S$ is fixed by gaugino condensation.

One might now be interested in the analysis of the mechanism of
supersymmetry breakdown and the properties of the soft terms.
The sources of supersymmetry breakdown  are vacuum expectation values
(vevs) of the auxiliary components
of the moduli superfields: $F_S$, $F_T$ and $F_{Z_i}$, and also
the auxiliary component of the 4D supergravity multiplet. In the KKLT
scheme, however, all but one of the
moduli have been integrated out. Therefore meaningful statements can
only be made for $F_T$ in the D3/7 case
(or $F_S$ in the heterotic case). The information on the remaining
moduli is hidden in $W_0$.
For a full analysis of the nature of supersymmetry breakdown we would
have to go a step back to a more fundamental level
and ``integrate in'' the other moduli. Only then can we make
meaningful statements about the relation of the values of the various
auxiliary fields and check if the assumption about integrating out
$S$ was consistent.

\section{The two modulus case: $S$ and $T$}

The simplest setup  is offered by models without complex structure
moduli (CSM) $Z_i$.
Alternatively one might
consider models where the complex structure moduli have been
integrated out. Later, however, we shall see that
this mechanism of integrating out moduli might be problematic and
therefore we here explicitly assume
the absence of CSM. We shall treat the two cases separately.

\subsection{Type IIB case: D3/7}

We consider the D3/7 system with two moduli $T = t + i \tau$ and
$S = s + i \sigma$ with  the K\"ahler potential and  superpotential
given by:
\bea &
\label{e.dk}
K = - 3 \log (T + \ov T) - \log (S + \ov S)
\,,\nl
W = A + B S + C e^{- a T}
\eea
where $A$, $B$, $C$ and $a$ are constants. The exponent $a$ is real and
positive. For simplicity we choose parameters $A$, $B$ and $C$ to be
real (the results in the general case are unchanged).
We choose $B$ and $C$ in such a way that $B \cdot C <0$. Then  we may
restrict our analysis to the stationary point with vanishing axion
vevs $\t=\s=0$  (for positive $B\cdot C$ we can just shift $\tau$ by
$\pi/a$).

The supersymmetric stationary point of the potential can be found by
solving the equations $F_S = F_T = 0$, where
$F_X \equiv  -e^{K/2} (K'')^{-1}{}_X^Y D_Y W$,
$D_Y W \equiv ({\pa K  \over \pa Y} W  + {\pa W \over \pa Y})$.
It occurs for the values of $s$  and $t$ satisfying the constraints:
\beq
\label{e.dsm}
\ds
\frac{Ce^{-at}}{A} = - {3 \over  a t + 3}
\,,\qquad
\frac{Bs}{A}  =  { a t \over a t + 3} \,.
\eeq
These constraints can be solved for $s$ and $t$ with the help of
the Lambert W function but the above form is more useful for our
analysis.
In the following we will assume that the parameters of the superpotential are such that the supersymmetric stationary point at large positive values of $at$ and $s$ (of order $10$ or larger) does exist.

In order to determine the nature of the stationary points we have to
consider the second derivatives of the potential $V_0$ at those
points. The explicit calculation shows, that all mixed real-imaginary
derivatives vanish
$\frac{\pa^2 V_0}{\pa t \pa\tau}=\frac{\pa^2 V_0}{\pa t \pa\sigma}
= \frac{\pa^2 V_0}{\pa s \pa\tau}=\frac{\pa^2 V_0}{\pa s \pa\sigma}=0$
and the second derivative  matrix
 splits into two $2\times 2$ matrices, $V_{0M}''$ and $V_{0A}''$.
After inserting the relations \erefn{dsm}
we find that all diagonal entries in these matrices are
positive. Nevertheless both the moduli and the axionic mass matrix has
one positive and one negative eigenvalue.
The instability can be easily seen by   calculating the determinant of
these matrices
\bea
&& {\rm det} V_{0M}''
= - \frac{3 B^4}{32 M_p^4 s^4 t^8} (4a^2 t^2 + 13 at + 10)
\,,\nn &&
{\rm det} V_{0A}''
= - \frac{3 B^4}{32 M_p^4 s^4 t^8} a t (4a t + 3 )
\eea
which are negative  for all $at>0 $. We conclude that the
supersymmetric point is a saddle point with  instabilities along the
moduli and axionic directions. One can also check that  other,
non-supersymmetric extrema of $V_0$ are also saddle
points. Furthermore, there exist directions in the $(t,\tau,\sigma)$
field space for which $\lim_{s\to 0^+}V_0=-\infty$ and the potential
is unbounded from below. This limit is, of course, outside the
validity of the perturbative supergravity approximation. Still we can
conclude that within the region of large S and large T there are no
local minima. All the stationary points are saddle points.

This shows that in this case the mechanism of KKLT faces problems. The
difference can be explained in the following way. KKLT first
considered the case with $C=0$ and a fixed dilaton. Then they
integrated out the dilaton before they included the gaugino condensate
(assuming implicitly that $m_s\,,m_\sigma\gg m_t\,,m_\tau$).
The mass of the dilaton, however, still depends on the $T$-modulus and
therefore this procedure is not necessarily justified. Our analysis
keeps  $S$ as well as $T$ after including the gaugino condensate and shows
that the assumption of KKLT is not justified at this level, as the
masses of $S$ and $T$ are comparable.

In principle, of course, one might say that even in the absence of
(supersymmetric) AdS minima one might arrive at a stable situation
after uplifting. We study a quite general lifting potential of the
form:
\beq
\label{e.hlp}
\Delta V = {D \over (T + \bar T)^{n_t}(S + \bar S)^{n_s}}
\eeq
and consider the potential $V = V_0 + \Delta V$. We assume that the
exponents $n_t$ and $n_s$ are positive or zero integers.
We look for solutions of the equations
${\pa V \over \pa T} = {\pa V  \over \pa S}= 0$
 with the parameter $D$ fine-tuned such that $V = 0$ (or a small
positive value). Such analysis shows that the situation in the
D3/7  cases remains unstable. We do not find any stable uplifted vacua
in the limit of large $at$. Both moduli and axionic instabilities
persist, independently of the exponents in the lifting potential. In
some cases there exist stable vacua at the small value of the moduli,
$at \sim 1$, but in this region the perturbative approximation of the
potential might not be reliable.

\subsection{Heterotic case: H}

We perform the similar analysis in the heterotic case. Here we take
\bea &
\label{e.hk}
K = - 3 \log (T + \ov T) - \log (S + \ov S)
\,,\nl
W = A + B T + C e^{- a S}  \,.
\eea
The supersymmetric stationary points exist if the parameters satisfy
the constraints: $C \cdot A > 0$, $B \cdot A < 0$ and $|C| \gg |A|$.
They occur for the values of $s$ and $t$ satisfying the conditions:
\beq
\label{e.hsv}
\frac{Ce^{-as}}{A} = {1 \over a s -1} \,,\qquad
\frac{Bt}{A}  = - {3 a s \over a s - 1} \,.
\eeq

Again the second derivative matrix at the stationary point  splits
into two $2\times 2$ blocks with all diagonal entries positive.
This time the moduli matrix $V_M''$ has two positive eigenvalues for
$a s >2$. But the instability is present in the axionic sector - the
determinant of $V_A''$ is negative for $as>0$.
We conclude that the supersymmetric point is a saddle point with an
instability along the axionic direction. One can also check that
other, non-supersymmetric extrema of $V_0$ are also saddle points.
Therefore for the heterotic system as well we do not obtain local
minima in the region of validity of our approximation. The behavior
of the limiting values of the potential is similar to that in the D3/7
case. There are directions in the moduli space for which the potential
is unbounded from below for $s\to 0^+$.

Note, however, that the situation is somewhat different than for the
D3/7 systems which have both moduli and axionic instabilities at the
supersymmetric point. Moreover, the character of the axionic mass
matrix can be changed by a small perturbation of the vacuum solution
in \eref{hsv}. Whether the stationary point is a minimum or a saddle
point depends on the  determinant of this matrix, which vanishes in
the leading order in $1/(as)$. The unstable character of the
supersymmetric  stationary point is determined only by the subleading terms.

As we have seen, the supersymmetric stationary points in the heterotic
case are only marginally unstable. Thus small corrections to the K\"ahler
potential might lead to stable minima, while this is not possible in
the D3/7 case.
In fact it can be shown that any correction to $K$ making
$(\partial_T K)^2/\partial_T^2 K> 3$ can lead to a stable supersymmetric AdS minimum.
In this regard, an interesting possibility is to have the higher order sigma model correction
yielding $K=-\ln\left[
(T+\bar{T})^3+E\right]$
where $E$ is a positive constant depending on the topological data of the underlying
compact manifold \cite{kaehler-correction}. We have analyzed this case as well as other modifications of
the K\"ahler potential which will be presented in a future publication.

The lifting potential can also provide for a small correction to the
vacuum solution and, under certain conditions, the stationary points
could become stable minima.
We add the lifting potential of \eref{hlp} and look for vacuum
solutions in the large $as$ limit and with vanishing cosmological
constant.
For $n_t\ne1$ the solution for which $C\exp(-at)/A$ behaves as
$1/(as)$ (similarly as the supersymmetric solution) reads:
\beq
\label{e.lsv}
\frac{Ce^{-as}}{A}  = {(n_t-4) \over 4 (n_t-1)}  {1 \over  a s}
+ \co\left(\frac{1}{(as)^2}\right)
\,,\qquad
\frac{Bt}{A}  = - {3 (n_t-2) \over 2 (n_t-1)}
+  \co\left(\frac{1}{a s}\right) \,.
\eeq
Only for $n_t = 0$ the above solution coincides with the
supersymmetric solution (\ref{e.hsv}) in the large $as$ limit.
For $n_t > 1$ the lifted solution is shifted with respect to the
supersymmetric solution already at the leading order in $1/(as)$. The
moduli directions stay stable for $n_t < 1$. For $n_t > 1$ the moduli
potential develops a saddle point instability (the case $n_t=1$ is
more complicated and will be analyzed fully elsewhere).
Calculating the determinant of the  axionic mass matrix
we find that for $n_t=0$ the axionic direction is also stable as
long as $n_s \geq 1$.

\section{Inclusion of the complex structure moduli}

\subsection{Type IIB case: D3/7}

We now include CSM $Z_i$ in the analysis. That is we study the system
described by
\bea &
\ds
K = - 3 \log (T + \ov T) - \log (S + \ov S)
+ \tilde K(Z_i,\bar{Z}_i)\,,
\nl\ds
W = W_{\rm flux}+Ce^{-a T}= A(Z_i) + B(Z_i) S + C e^{- a T}\,.
\eea
In this case $W_{ij}=\frac{\partial^2 W_{\rm flux}}{\partial Z_i\partial Z_j}$
and $W_{iS}=\frac{\partial^2 W_{\rm flux}}{\partial Z_i\partial S}$ at supersymmetric vacuum
are non-vanishing in general \cite{coqu}.
If $W_{ij} > W_{iS}$, $Z_i$ can be integrated out first, leaving an effective superpotential
for $S$.
In the following, we consider such situation that $Z_i$ are heavy enough to be integrated out,
while $m_S$ has an arbitrary value between $m_{Z_i}$ and $m_{3/2}$.

The resulting effective superpotential $W^S_{\rm eff}[S]$ depends on the  precise form of  $A(Z_i)$ and $B(Z_i)$.
 We parametrize our ignorance  by
assuming that integrating out $Z_i$ leaves
some general function of $S$ in the effective
superpotential. Of course one might have assumed that the case
discussed in the previous section, with $W^S_{\rm eff}=A+BS$, has been
obtained after integrating out CSM. We have already learned that the
conclusions of KKLT do not hold in this specific case. 
We will study the equations of
motion for general $W^S_{\rm eff}[S]$ and determine the conditions
that $W^S_{\rm eff}$ must satisfy in order to arrive at a stable
vacuum. In the next section we also study the dependence of soft
breaking terms on $W^S_{\rm eff}$.
Thus, we consider the supergravity setup defined by:
\bea &
\label{e.dgk}
\ds
K = - 3 \log (T + \ov T) - \log (S + \ov S)
\,,\nl\ds
W = W^S_{\rm eff}[S] + C e^{- a T} \,.
\eea
Again, for simplicity we assume that $C$ and $W_{\rm eff}[s]$ are
real. The supersymmetric stationary points occur for $s$  and $t$
satisfying the constraints:
\beq
\label{e.dgsv}
\ds
\frac{Ce^{-at}}{W^S_{\rm eff}}
= - \frac{3}{3+2at}
\,,\qquad
\frac{s{W^S_{\rm eff}}'}{W^S_{\rm eff}}
= \frac{at}{3+2at}\,.
\eeq
To determine the character of the stationary point we must study the
second derivatives.  The relevant parameter here is $\gamma$ defined
as:
\beq
\gamma \equiv
{s\, W^S_{\rm eff}{}'' \over {W^S_{\rm eff}}'} 
\eeq
with $s$ satisfying the constraints \erefn{dgsv}. The second
derivative matrices after inserting the  relation \erefn{dgsv} read:
\beq
\label{e.dgmm}
V_{0M}'' =  \frac{{{W^S_{\rm eff}}'}^2}{8 M_p^2 s t^5}
\left [ \ba{cc}
 3 s^2 (4a^2 t^2 + 10 at + 7)  &
3  s t ( 2 a t +3 - 2\gamma)  \\
 3 s t ( 2 a t +3 - 2\gamma)  &
 t^2 ( 1 - 2\gamma + 4\gamma^2)
\ea\right ]
\eeq
for the moduli ($t,s$) and
\beq
\label{e.dgam}
V_{0A}'' =  \frac{{{W^S_{\rm eff}}'}^2}{8 M_p^2 s t^5}
\left [ \ba{cc}
 3 s^2 (4a^2 t^2 + 6 at + 3)
& 3 s t ( 2 a t +1 - 2\gamma)  \\
 3 s t ( 2 a t +1 - 2\gamma)
&  t^2 ( 1 + 2\gamma + 4\gamma^2)
\ea\right ]
\eeq
for the axions ($\tau,\sigma$). All diagonal entries are manifestly
positive. Whether a stationary point is a minimum or a saddle point depends
on the signs of the determinants. The general conditions for
positivity of the determinants are rather complicated. The case
$\gamma=0$ was studied in the previous section and we found
instabilities in both axionic and moduli direction. Therefore the KKLT
setup with a simple linear superpotential $W^S_{\rm eff}[S] = A + BS$
is inconsistent. Stability can be achieved only when the second
derivative of the $S$-dependent effective  superpotential at the
minimum $s^2 {W^S_{\rm eff}}''$  is at least comparable to $W^S_{\rm eff}$ and
$s {W^S_{\rm eff}}'$, that is for $\gamma$ of order one or bigger. In
particular, in the large $(a \, t)$ limit the stability condition is
very simple:
\beq
|\gamma| > 1
\eeq
For $\gamma \gg 1$ and $\gamma \gg a t$   the  $S$ modulus becomes
much heavier than $T$ and is decoupled.
 This is the KKLT limit. 
For $\gamma \to \infty$ we should recover all the results of KKLT. 
The actual magnitude of $\gamma$ depends on the  flux compactification model under
 consideration. Note that, by vacuum equations \erefn{dgsv}, large $a t$  of order $10$ implies 
$C \gg s  W^S_{\rm eff}{}'$, thus   $W^S_{\rm eff}{}'$ must be suppressed with respect to the string scale, 
such that $s  W^S_{\rm eff}{}'/C < 10^{-3}$. 
This can be achieved if  the flux parameters are appropriately fine-tuned or if there exist some hierarchy of parameters in  $W^S_{\rm eff}$. In the former case one would expect $\gamma \gg 1$ 
(unless another fine-tuning makes   $W^S_{\rm eff}{}''$ suppressed as well),
 in the latter case $\gamma \sim 1$ can be  natural.

In order to arrive at a vacuum with a vanishing (or small positive)
cosmological constant we include the lifting potential \eref{hlp}.
 The analysis is quite complicated and we shall restrict ourselves to
consider only the large $(a\,t)$ limit (which includes  the KKLT limit when $\gamma \gg at$).
In the former case, after  eliminating $D$ and $t$ we find the vev of
$s$ is given by the solution of the equation
\beq
(n_s -1) {W^S_{\rm eff}}^2
+2 s (n_s + 1) W^S_{\rm eff} {W^S_{\rm eff}}'
- 2 n_s s^2 {{W^S_{\rm eff}}'}^2
+ 2 s^2  W^S_{\rm eff} {W^S_{\rm eff}}''
- 4 s^3 {W^S_{\rm eff}}' {W^S_{\rm eff}}''
   =0
\eeq
 This can be easily solved for $W^S_{\rm eff}$ to give the vacuum
 solution in a form similar to \eref{dgsv}.
The most compact form  is obtained for  $n_s = 1$ and in the following
we restrict to this case (for $n_s \neq 1$ there is no qualitative
difference).
In this special case we get:
\beq
\label{e.dglvl}\ds
\frac{Ce^{-at}}{W^S_{\rm eff}}  = -\frac{3}{2at}
+ \co\left(\frac{1}{(at)^2}\right)
\,,\qquad
\frac{s{W^S_{\rm eff}}'}{W^S_{\rm eff}}
= \frac{2+\gamma}{1+2\gamma}  +  \co\left(\frac{1}{a t}\right) \,.
\eeq
The stability condition is somewhat more complicated (it depends also
on ${W^S_{\rm eff}}'''$) but generically it requires
$\gamma \simgt 1$ as in the \ads \, vacuum.
We see that, in general, even in the large $(at)$ limit the lifted
vacuum is shifted with respect to the position of the supersymmetric
\ads \, vacuum, see \eref{dgsv}. Only for   $ \gamma \gg 1 $ the position
of the lifted vacuum coincides with that of the supersymmetric vacuum
(in fact eqs. (\ref{e.dglvl}) hold for arbitrary $n_s$ then). The lifted vacuum is always stable in this limit.

\subsection{Heterotic case: H}

One can do the analogous analysis for the heterotic case defined by
\bea &
\label{e.hgk}
\ds
K = - 3 \log (T + \ov T) - \log (S + \ov S)
\,,\nl\ds
W = W^T_{\rm eff}[T] + C e^{- a S}
\eea
where $W^T_{\rm eff}[T]$ represents the $T$-dependent part of the
superpotential after integrating out the CSM. The supersymmetry
preserving configurations satisfy the following conditions for the
moduli $t$ and $s$:
\beq
\label{e.hgsv}
\ds
\frac{Ce^{-as}}{W^T_{\rm eff}}
= - \frac{1}{1+2as}
\,,\qquad
\frac{t\,{W^T_{\rm eff}}'}{W^T_{\rm eff}}
= \frac{3as}{1+2as}\,.
\eeq
The character of those stationary points depends on the second
derivative matrices which are given by
\beq
\label{e.hgmm}
V_{0M}'' =  \frac{{{W^T_{\rm eff}}'}^2}{ 72 M_p^2 s^3 t^3}
\left [ \ba{cc}
 3 s^2 (4\eta^2-10\eta+7)  &
3  s t ( 2 a s +3 - 2\eta)  \\
 3 s t ( 2 a s +3 - 2\eta)  &
 t^2 ( 4a^2s^2+2as+1)
\ea\right ]\,,
\eeq
\beq
\label{e.hgam}
V_{0A}'' =  \frac{{{W^T_{\rm eff}}'}^2}{ 72 M_p^2 s^3 t^3}
\left [ \ba{cc}
 3 s^2 (4\eta^2-6\eta+3)
& 3 s t ( 2 a s +1 - 2\eta)  \\
 3 s t ( 2 a s +1 - 2\eta)
&  t^2 ( 4a^2s^2 -2as +1)
\ea\right ]
\eeq
where $t$ and $s$ satisfy conditions \erefn{hgsv} and the parameter
$\eta$ is defined by

\beq
\eta \equiv
{ t W^T_{\rm eff}{}'' \over {W^T_{\rm eff}}'} \,.
\eeq
After calculating determinants of the above matrices we find that in
the leading order in $1/(as)$ the supersymmetric stationary points
\erefn{hgsv} are stable minima for
\beq
|\eta-1| > 1 \,.
\eeq
One can see that in the absence of CSM the
heterotic model with $\eta=0$ is just
at the border of the (in)stability region.

The supersymmetric stationary points (stable or not) satisfying
 \eref{hgsv} have negative vacuum energy. We can lift them to
zero or small positive value by adding the lifting potential of the
form \erefn{hlp}. The conditions for the minima with vanishing energy
are in general quite complicated. In the leading order in $1/(as)$
the values of the moduli $s$ and $t$ at such minima (for $n_t\ne3$)
 satisfy:
\beq
\label{e.hglvl}\ds
\frac{Ce^{-as}}{W^T_{\rm eff}}  = -\frac{1}{2as}
+ \co\left(\frac{1}{(as)^2}\right)
\,,\qquad
\frac{t{W^T_{\rm eff}}'}{W^T_{\rm eff}}
= \frac{3(2-n_t-\eta)}{4-n_t-2\eta}  +  \co\left(\frac{1}{a s}\right) \,.
\eeq
These conditions coincide in leading order with the conditions
before lifting given by \eref{hgsv} only for $n_t=0$ or in the large
$\eta$ limit.

\section{Soft supersymmetry breaking terms}

In this section, we discuss the soft SUSY breaking terms in various setups we have considered
so far. We will concentrate on soft terms induced by the auxiliary components
of $S$, $T$ and the 4D supergravity multiplet under the assumption that
CSM are heavy enough to be ignored.
We also ignore the soft terms that might possibly be 
induced by the direct couplings
between the visible sector and the uplifting sector.

Before presenting the results for the vevs of the auxilary components in KKLT compactifactions 
let us summarize the important facts about supersymmetry breaking mediation in supergravity \cite{tree-soft,AMSB,loop-soft}. 
The auxiliary component of 4D supergravity multiplet can be parameterized by
a chiral compensator $\phi=1+\theta^2\hat{F}_\phi$ whose $F$-component
is given by
\beq
\hat{F}_\phi= m_{3/2}+\frac{1}{3}F_A\partial_A K =m_{3/2}-\frac{1}{3}\hat{F}_S-\hat{F}_T \, , 
\eeq
where $\hat{F}_A=F_A/(A+\bar{A})$.
The auxiliary component $\hat{F}_\phi$ can induce soft terms (other than $B$) only at loop-level by the mechanism of anomaly mediation \cite{AMSB}. As we will see, in some interesting limits of flux compactifications,
$\hat{F_\phi}$ is bigger than
$\hat{F}_{S,T}$ by one or two orders of
magnitudes, e.g.
$\hat{F}_\phi={\cal O}\,(8\pi^2 \hat{F}_{S,T})$.
In this case, the loop-induced soft terms associated with $\hat{F}_\phi$ can be
equally important as the dilaton and/or modulus-mediated soft terms at tree-level.

To accomodate the loop-induced soft terms,
let us consider the superspace lagrangian of the visible
fields which includes quantum corrections at the compactification
scale $M_c$:
\bea
\label{superlagrangian}
{\cal L}_{\rm visible}&=&
\int d^4\theta \,\left[Y_i\bar{Q}_iQ_i +\frac{1}{16}
\left(G_a W^{a\alpha}\frac{D^2}{\partial^2}W^a_\alpha
+{\rm h.c}\right)\,\right]
\nonumber \\
&+& \left[\,\int d^2\theta\,
\frac{1}{6}\,\lambda_{ijk}Q_iQ_jQ_k+{\rm h.c}\,\right],
\eea
where the holomorphic Yukawa couplings $\lambda_{ijk}$ are assumed to be
moduli-independent constants.
At tree level, $G_a$ corresponds to the holomorphic gauge kinetic function
$f_a$.
In 4D supergravity, the superspace lagrangian of a K\"ahler potential
$K$ is given by $\int d^4\theta [-3e^{-K/3}]$,  thus
$$
Y_i= (S+\bar{S})^{1/3}(T+\bar{T}){\cal Z}_i,
$$
for $K=-\ln(S+\bar{S})-3\ln(T+\bar{T})+
{\cal Z}_i \bar{Q}_iQ_i$.
From (\ref{kaehler-IIA}), (\ref{gauge-IIA}) and (\ref{kaehler-het}), one easily finds the
tree-level expressions of  $G_a$ and $Y_i$ for the D3,D7 and heterotic matter/gauge fields:
\bea
\label{tree}
 G_3^{(0)}=G_H^{(0)}=S,
&&\,
G_7^{(0)}=T,
\nonumber \\
Y^{(0)}_3= Y^{(0)}_H= (S+\bar{S})^{1/3},
&& \,
Y^{(0)}_7 =(S+\bar{S})^{-2/3}(T+\bar{T}),
\eea
where the superscript $(0)$ means the tree-level result.

The physical gauge and Yukawa couplings renormalized at 
$M_c$ are given by
\beq
\frac{1}{g_a^2}= {\rm Re}(G_a),
\quad
y_{ijk}= \frac{\lambda_{ijk}}{\sqrt{Y_iY_jY_k}}.
\eeq
Let us define the canonically normalized
gaugino masses, $A$-parameters and scalar masses
as
\beq
\-\frac{1}{2}M_a\lambda^a\lambda_a -
{1 \over 2} {m_i}^2 \left|\tilde{Q}_i\right|^2 
- \frac{1}{6}A_{ijk}y_{ijk}\tilde{Q}_i\tilde{Q}_j\tilde{Q}_k+{\rm h.c.}
\eeq
Generically these soft masses are given by
\bea
M_a &=& c_a^S\hat{F}_S+c_a^T\hat{F}_T+c_a^\phi\hat{F}_\phi,
\nonumber \\
A_{ijk}&=& a^S_{ijk}\hat{F}_S+a^T_{ijk}\hat{F}_T+a^\phi_{ijk}\hat{F}_\phi,
\nonumber \\
m_i^2&=& h^{S\bar{S}}_i\left|\hat{F}_S\right|^2
+h^{T\bar{T}}_i\left|\hat{F}_T\right|^2+
\left(h^{S\bar{T}}\hat{F}_S\bar{\hat{F}}_T+{\rm h.c}\right)
\nonumber \\
&+& h^{\phi\bar{\phi}}_i\left|\hat{F}_\phi\right|^2
+\left(h^{S\bar{\phi}}_i\hat{F}_S\bar{\hat{F}}_\phi+
h^{T\bar{\phi}}_i\hat{F}_T\bar{\hat{F}}_\phi+{\rm h.c}\right).
\eea
In our case including the limit with $\hat{F}_\phi={\cal O}(8\pi^2 \hat{F}_{S,T})$,
the dominant parts of soft masses at $M_c$ can be determined
by the tree-level values of
$c_a^{A}$, $a_{ijk}^{A}$ and $h^{A\bar{B}}_i$ ($A,B=S,T$),
the one-loop values of $c_a^\phi$, $a_{ijk}^\phi$ and $h^{A\bar{\phi}}$,
and also the two-loop values of $h_i^{\phi\bar{\phi}}$.
Inserting (\ref{tree}) into the superspace lagrangian (\ref{superlagrangian}), we find
the following tree-level results for the D3,D7 and heterotic matter/gauge fields
\cite{tree-soft}:
\bea
&&c_{3}^S=c_H^S=1, \quad c_{3}^T=c_H^T=0,\quad  c_7^S=0,\quad
c_7^T=1,
\nonumber \\
&&a^A_{ijk}=\kappa^A_i+\kappa^A_j+\kappa^A_k,
\quad h^{A\bar{A}}_i=\kappa^A_i,
\quad
h^{A\bar{B}}_i=0 \,\,\,(A\neq B),
\eea
where
\bea
\kappa^S_{3}=\kappa^S_H=\frac{1}{3},\quad
\kappa^T_{3}=\kappa^T_H=0,\quad
\kappa^S_7=-\frac{2}{3},\quad
\kappa^T_7=1.
\nonumber
\eea
The one-loop values of $c_a^\phi$ and $a_{ijk}^\phi$ and the
two-loop values of $h^{\phi\bar{\phi}}_i$ correspond to the well-known anomaly-mediated
soft terms \cite{AMSB}:
$c_a^\phi$ corresponds to the one-loop beta function coefficient
$\frac{dg_a}{d\ln\mu}=c_a^\phi g_a^3$, while
$a^\phi_{ijk}=-\frac{1}{2}(\gamma_i+\gamma_j+\gamma_j)$ and $h^{\phi\bar{\phi}}_i=
-\frac{1}{4}\frac{d\gamma_i}{d\ln\mu}$
where $\gamma_i=\frac{d\ln Y_i}{d\ln\mu}$ is the anomalous dimension of $Q_i$.
These anomaly-mediated soft masses can be most easily computed by
including the $\phi$-dependent one-loop corrections of  $G_a$ and $Y_i$ in the
superspace lagrangian (\ref{superlagrangian})
\cite{AMSB,loop-soft}:
\bea
{\rm Re}(G_a)&=&{\rm Re}(f_a) -\frac{1}{16\pi^2}\left(3T_a({\rm Adj})-\sum_i T_a(Q_i)\right)\ln\left(\frac{
\phi\bar{\phi}}{\mu^2}\right)+...,
\\
\ln \left(Y_i\right)&=& \ln\left(Y_i^{(0)}\right)-\frac{1}{32\pi^2}\left(
4\sum_a\frac{T_a(Q_i)}{{\rm Re}(f_a)}-\sum_{jk}\frac{|\lambda_{ijk}|^2}{Y^{(0)}_iY^{(0)}_jY^{(0)}_k}
\right)\ln\left(\frac{\phi\bar{\phi}}{\mu^2}\right)+...,
\nonumber \eea
where $T_a$ denotes the quadratic Casimir
and the ellipses stand for the $\phi$-independent (but generically moduli-dependent)
loop corrections which are not relevant for us.
These $\phi$-dependent parts of $G_a$ and $Y_i$  determine
$h^{A\bar{\phi}}$ also as
\bea
\Delta m_i^2 &\equiv&  h^{S\bar{\phi}}\hat{F}_S\bar{\hat{F}}_\phi
+h^{T\bar{\phi}}\hat{F}_T\bar{\hat{F}}_\phi+{\rm h.c}
\nonumber \\
&=&\frac{1}{32\pi^2}\left(
\sum_{jk}|y_{ijk}|^2 A^{(0)}_{ijk} -
4\sum_a g_a^2 T_a(Q_i) M_a^{(0)}\right)\bar{\hat{F}}_\phi+{\rm h.c},
\eea
where $M_a^{(0)}=c_a^S\hat{F}_S+c_a^T\hat{F}_T$ and $A^{(0)}_{ijk}=a^S_{ijk}\hat{F}_S+
a^T_{ijk}\hat{F}_T$
are the tree-level gaugino masses and $A$-parameters.

 Let us now discuss the relative ratios between $\hat{F}_S$,
$\hat{F}_T$ and $\hat{F}_\phi$ that we found in KKLT compactifications.
For the D3/7 system in the limit $at\gg 1$ with $n_s =1$, we
find
\bea &
\ds
 m_{3/2} \approx  \frac{s {W^S_{\rm eff}}'}{2 M_p^2}\,
{1 + 2 \gamma \over 4 +2\gamma} \,  s^{-1/2} t^{-3/2}
\,,\nl
\ds \hat{F}_S  \approx
 - {3 \over 1 + 2 \gamma}\,  m_{3/2}
\,,\nl
\ds \hat{F}_T \approx
\frac{1}{at}\, {6-n_t+ \gamma(3+2n_t)
+ 2 \gamma^2 n_t \over (1 + 2\gamma)^2} \, m_{3/2} \,.
\eea
From (\ref{e.dglvl}), one easily finds $at\sim \ln (M_p/m_{3/2})$. 
If $m_{3/2}$ is of the order of the TeV scale then $at$ can be as large as $35$. 
In such case, $1/at$ is comparable to the one-loop suppression factor, $at={\cal O}(8\pi^2)$.
The above results show that $\hat{F}_S={\cal O}(m_{3/2})$
and $\hat{F}_T={\cal O}(m_{3/2}/at)$
for $\gamma \sim 1$. In this case, all soft terms on D3 brane and also
the scalar masses on D7 brane are
dominated  by the contributions from $\hat{F}_S$,
while the gaugino masses and $A$-parameters on D7 branes  receive equally important contributions from
$\hat{F}_T$ and $\hat{F}_\phi$.

On the other hand, in the KKLT limit of the D3/7 system in which $m_S\gg m_T$ and thus $|\gamma|\gg at$,
one always finds (for arbitrary $n_s$)
$\hat{F_T}={\cal O}(m_{3/2}/at)$ and $|\hat{F}_S|\ll |\hat{F}_T|$.
Thus in the KKLT limit with weak scale supersymmetry
in which $at\sim \ln(M_p/m_{3/2})={\cal O}(8\pi^2)$,
soft terms on D3 brane are dominated by anomaly-mediation, while those
on D7 brane are dominated by the equally important contributions
from $\hat{F}_T$ and
$\hat{F}_\phi$.

In the heterotic case without CSM, we were also
able to find a stable vacuum solution after including the lifting
potential. Here  the pattern of soft breaking terms is  different.
For $n_t = 0$ and $as\gg 1$,
 we find
\bea &
\ds
 m_{3/2} \approx  \frac{A}{2M_p^2}\,  s^{-1/2} t^{-3/2}
\,,\nl
\ds \hat{F}_S  \approx
 - {1 \over as}\, {3 n_s \over 2}\, m_{3/2}
\,,\nl
\ds \hat{F}_T \approx
 {1 \over as}\, {3 n_s \over 8}\,   m_{3/2} \,.
\eea
Again (\ref{e.lsv}) indicates $as\sim \ln(M_p/m_{3/2})$.
Although $\hat{F}_S\sim \hat{F}_T$,
the soft masses from $\hat{F}_T$
appear only at higher orders in either
string loop expansion or $\alpha'$-expansion due to the no-scale nature.
Thus, in this heterotic setup with weak scale supersymmetry,
soft masses are dominated  by the equally important contributions
from $\hat{F}_{S}$ and $\hat{F}_\phi$.

For the heterotic case with CSM and the
lifting potential with $n_s=1$,
supersymmetry breaking parameters for $as\sim \ln(M_p/m_{3/2})\gg 1$ are given by
\bea &
\ds
 m_{3/2} \approx  \frac{t {W^T_{\rm eff}}'}{12 M_p^2}\,
\frac{4-n_t-2\eta}{2-n_t-\eta} \,  s^{-1/2} t^{-3/2}
\,,\nl
\ds \hat{F}_S  \approx
 \frac{1}{as}\, \frac{3(2-n_t-\eta)}{4-n_t-2\eta}\, m_{3/2}
\,,\nl
\ds \hat{F}_T \approx
\frac{n_t}{4-n_t-2\eta}\, m_{3/2} \,.
\eea
For $\eta\sim 1$ and $n_t\neq 0$, we have
$\hat{F}_T={\cal O}(m_{3/2})$, while $\hat{F}_S={\cal O}(m_{3/2}/as)$.
In this case, all of $\hat{F}_S$, $\hat{F}_T$ and $\hat{F}_\phi$
give similar contributions to soft masses.
On the other hand, in the KKLT limit of heterotic set up in which
$m_T\gg m_S$ and thus $|\eta|\gg 1$, soft terms
are dominated by the equally important contributions from
$\hat{F}_S$ and $\hat{F}_\phi$.
This feature is very similar
to the D3/7 case if we
interchange $s$ with $t$ and $\gamma$ with $\eta$.

\section{Conclusions and outlook}
We have extended the KKLT approach to moduli stabilization by
explicitly including the dilaton (or $T$ in the heterotic model) and
the complex structure moduli into the effective supergravity
theory. The conditions on the effective superpotential for the dilaton
(or $T$) field have then been derived that allow for the existence of
 supersymmetric AdS vacua and a stable minimum after supersymmetry
breaking and uplifting of the scalar potential. These conditions
provide restrictions on the CSM and the flux configurations once
explicit string models are constructed. However, they admit a much
more general class of models than those in which the dilaton $S$
(K\"ahler modulus $T$ in the heterotic case) can be decoupled. In
consequence, the pattern of supersymmetry breaking as given in
equations (37-39) can be much richer
than $F_S \ll F_T \ll m_{3/2}$ 
(as in the  KKLT limit where anomaly mediation plays an important role)
and there are regions of parameter space in which  
tree-level mediation becomes dominant.  
Therefore a generalization of the scheme beyond the KKLT limit seems
to be required before definite conclusions can be drawn. A more
complete discussion of this situation together with the full
technical details will be the subject of a future publication
\cite{future}.

\vspace{5mm}
\noindent{\large\bf Acknowledgments}
\vspace{5mm}

K.C. and H.P.N. would like to thank Gary Shiu for interesting discussions. 
This work was partially supported by the EU 6th Framework Program
MRTN-CT-2004-503369 ``Quest for Unification'' and
MRTN-CT-2004-005104 ``ForcesUniverse''.
K.C. is supported by Korean KRF PBRG 2002-070-C00022.      
A.F. was partially supported by the Polish KBN grant 2 P03B 129 24
for years 2003-2005.
M.O. was partially supported  by the Polish KBN grant 2 P03B 001 25
for years 2003-2005.
S.P. would like to thank Hans Peter Nilles (Bonn) and Jan Louis
(Hamburg) for their hospitality. His visits to the University of Bonn
and University of Hamburg were possible thanks to a ``Forschungs-Preis'' of
the A. von Humboldt Foundation. A.F. would like to thank 
DESY (Hamburg) for hospitality and acknowledge 
A. von Humboldt Foundation Research Fellowship 
that made possible his stay at DESY. 
K.C. and H.P.N. acknowledge the support of the Aspen Center for Physics
during the early stages of this work.

\vspace{5mm}
\noindent{\large\bf Note added}
\vspace{5mm}

While completing this work we became aware of a paper 
by Kallosh and Linde \cite{Kallosh}
where the simplest KKLT version has been modified. Also this version
is formulated in the KKLT limit ($|\gamma|\gg at$) and should be generalized
along the lines discussed above before the full picture of the
pattern of supersymmetry breakdown can be analyzed.


\end{document}